\documentclass[12pt,a4paper]{JHEP3}

\usepackage{picinpar}
\usepackage{epsfig}
\usepackage{amsmath}
\usepackage{amssymb}
\usepackage{cite}
\usepackage{graphicx}

\newcommand{\beqs}{\begin{equation*}}

\newcommand{\beq}{\begin{equation}}

\newcommand{\eeqs}{\end{equation*}}

\newcommand{\eeq}{\end{equation}}

\newcommand{\beqas}{\begin{eqnarray*}}

\newcommand{\bea}{\begin{eqnarray}}

\newcommand{\eeqas}{\end{eqnarray*}}

\newcommand{\eea}{\end{eqnarray}}

\newcommand{\nn}{\nonumber}

\newcommand{\eq}[2]{\begin{equation} #1 \label{#2} \end{equation}}

\newcommand{\eqa}[2]{\begin{eqnarray} #1 \label{#2} \end{eqnarray}}

\newcommand{\de}{\delta}

\newcommand{\om}{\omega}

\newcommand{\blist}{\begin{itemize}}

\newcommand{\elist}{\end{itemize}}

\providecommand{\href}[2]{#2}

\DeclareFontFamily{OT1}{rsfs}{}

\DeclareFontShape{OT1}{rsfs}{m}{n}{ <-7> rsfs5 <7-10> rsfs7 <10->rsfs10}{} 

\DeclareMathAlphabet{\mycal}{OT1}{rsfs}{m}{n}

\newcommand{\scri}{{\mycal I}}

\DeclareMathOperator{\extdm}{d}

\newcommand{\extd}{\extdm \!}

\title{Flat/AdS boundary conditions in three dimensional conformal gravity}

\author{Hamid R.~Afshar\\
           Institute for Theoretical Physics, 
           Vienna University of Technology,\\
           Wiedner Hauptstr. 8--10/136,
           A-1040 Vienna, Austria. Europe\\
           Email: \email{afshar@hep.itp.tuwien.ac.at}}

\abstract{
We present the asymptotic analysis of 3D conformal gravity as a SO(3,2) Chern-Simons gauge theory with Minkowskian (flat) and AdS boundary conditions.
We further extend these boundary conditions to the case where the Weyl mode and the partial massless mode are allowed to fluctuate. 
The latter leads to loosing one copy of the Virasoro algebra and the former to a $\hat u(1)_k$ current extension of the asymptotic symmetry algebra and shifting the Virasoro central charge by one. 
We also give a pedagogical canonical and asymptotic analysis of 3D pure gravity as an ISO(2,1) Chern-Simons gauge theory with flat boundary conditions. 

}


 \preprint{TUW--13--10}

\begin{document}

\section{Introduction}
The Chern-Simons (CS) action in three dimensions has attracted much attention from different angles,
as a pure topological 3D gauge theory \cite{Witten:1988ze} or as a pure 3D gravity theory \cite{Deser:1981wh}. 
The observation of Ach\'{u}carro and Townsend in \cite{Achucarro:1986vz} and 
Witten in \cite{Witten:1988hc} that general relativity in three dimensions with and without cosmological constant
is equivalent to a Chern-Simons gauge theory with a proper gauge group, made these two view points even closer.
Attentions to three dimensional gravity which is empty in terms of local degrees of freedom were boosted by the discovery of boundary degrees 
of freedom \cite{Brown:1986nw} and topological degrees of freedom namely BTZ black holes \cite{Banados:1992wn,Banados:1992gq}. 
Microstates of these locally-AdS black holes were holographically counted by Strominger in \cite{Strominger:1997eq}.
The very origin of these microstates however, needs a full holographic description which is still missing \cite{Witten:2007kt,Maloney:2007ud,Castro:2011zq}.

One approach is to modify the bulk theory by adding more gauge symmetries which would lead to new boundary states. 
In this respect, the Chern-Simons formulation of the bulk gravity theory is very much privileged as it holds the action 
in the bulk still topological. First steps in this regard were made by Horne and Witten \cite{Horne:1988jf} by studying 
conformal gravity and Blencowe by studying higher spin gravity \cite{Blencowe:1988gj}. However the asymptotic symmetry
analysis \`{a} la Brown-Henneaux of these extensions -- with AdS boundary conditions -- were not surveyed until very 
recently in \cite{Afshar:2011yh,Afshar:2011qw} and \cite{Henneaux:2010xg,Campoleoni:2010zq} respectively.
Holographic principle is however not restricted to only AdS boundary conditions but extends to non-AdS ones \cite{Afshar:2012nk},
most importantly Minkowskian (flat) boundary conditions seems very suggestive.

In this paper we will present the asymptotic symmetry analysis of three dimensional pure gravity (without cosmological constant) and conformal gravity 
in Chern-Simons formulation by introducing suitable flat boundary conditions.
For generalization of this analysis to the higher spin case see \cite{Afshar:2013vka}. 
We also discuss the AdS boundary conditions in the absence  and presence of the partial massless mode.

This paper is organized as follows.
In section \ref{se:2} we review some general features of CS theories including its canonical analysis. 
In section \ref{se:3} we review the CS formulation of 3D pure  gravity and propose a consistent set of flat boundary conditions. 
We then construct boundary charges and derive the asymptotic symmetry algebra (ASA).
In section \ref{se:4} we do the same analysis for 3D conformal gravity with two consistent sets of boundary conditions, 
namely AdS and flat and find the ASA and discuss its representation. 
In section \ref{se:5} we summarize our results.

\section{Chern-Simons gauge theory}\label{se:2}
In this section, for sake of self-containment and fixing the notation, we review some known features of CS theory on a spacetime with boundary.
For an early reference see \cite{Banados:1994tn}.

We start with the following Chern-Simons action on a manifold with the topology $\mathcal{M} =\Sigma\times\mathbb{R}$, and assume that $\partial\Sigma=S^1$,
\eqa{
I_{\textrm{\tiny CS}}&=&\frac{k}{4\pi}\int_\mathcal{M}  \left\langle A\wedge dA+\tfrac{2}{3} A\wedge A\wedge A\right\rangle\,,
}{eq:ch1}
where $A$'s are Lie algebra-valued 1-forms, $A=A_{\mu}dx^{\mu}$, with the curvature two form, $F=dA+A\wedge A$. 
Thus if we choose a basis of the Lie algebra, and write $A=A^aT_a$, then $\langle T_a\,,\,T_b\rangle=g_{ab}$, plays the role of a non-degenerate invariant bilinear form
on the Lie algebra. We can write \eqref{eq:ch1} in components as,
\eq{
I_{\textrm{\tiny CS}}=\frac{k}{4\pi}\int_\mathcal{M} d^3x\,\epsilon^{\mu\nu\lambda}g_{ab}\left(A^a{}_\mu \partial_\nu A^b{}_\lambda+\tfrac{1}{3} f^a{}_{cd}A^c{}_\mu A^d{}_\nu A^b{}_\lambda\right)\,,}{eq:ch9}
where $f^a{}_{bc}$ are structure constants of the underlying Lie algebra.
\subsection{Field equations and gauge symmetries}
In order to have  a well-defined variational principle we should impose boundary conditions such that the variation of the action under generic variations of fields becomes zero on-shell.
Varying the total action \eqref{eq:ch1} for an arbitrary deformation in the phase space we have,
\bea\label{varprin}
\delta I|_{F=0}=-\frac{k}{4\pi}\int_{\partial \mathcal{M}}\langle A\wedge\delta A\rangle\,.
\eea
In order to have a well defined variational principle we should have this boundary contribution zero,
this restricts our boundary conditions. Adding additional boundary terms will surely change the 
variational principle \cite{Gary:2012ms,Afshar:2012nk}.

Let us check the gauge invariance of Chern-Simons theory (off-shell) under a general gauge transformation, $A \rightarrow g^{-1}( A+d)g$.
For any infinitesimal gauge transformation connected to identity $g\simeq 1+\varepsilon^a T_a$, the gauge field 
transforms;
\bea\label{gtfgen}
\delta_\varepsilon A^a{}_\mu=\partial_\mu{\varepsilon^a}+f^a{}_{bc}A^b{}_\mu\varepsilon^c\,,
\eea
and so does the action, 
\bea\label{deltI}
\delta_\varepsilon I= -\frac{k}{4\pi}\int_{\partial\mathcal{M}}\langle A\wedge  d\varepsilon\rangle\,.
\eea
The theory does not remain invariant under \eqref{gtfgen}, in fact
its gauge invariance depends on our choice of boundary conditions and sets of transformations we use.

\subsection{Gauge generators and boundary charges}
Using the $2+1$ decomposition, 
the Lagrangian density reads \cite{Blagojevic:2002du},
\eq{
\mathcal{L}_{\textrm{\tiny CS}}=\frac{k}{4\pi}\,\epsilon^{ij} g_{ab}\left(\dot{A}^a{}_i {A^b}_j+{A^a}_0{F^b}_{ij}+\partial_j(A^a{}_iA^b{}_0)\right).
}{eq:ch5}
Introducing  the canonical momenta ${\pi_a}^\mu\equiv\partial\mathcal{L}/\partial \dot{A}^a{}_\mu$ corresponding to the canonical
variables ${A^a}_\mu$, we find  primary constraints,
\eq{
{\phi_a}^0:={\pi_a}^0\approx0\,,\qquad{\phi_a}^i:={\pi_a}^i-\frac{k}{4\pi}\,\epsilon^{ij}g_{ab}{A^b}_j\approx0\,,
}{eq:ch6}
and the canonical Hamiltonian density is,
\eq{
\mathcal{H}=\pi_a{}^\mu \dot{A}^a{}_\mu-\mathcal{L}=-\frac{k}{4\pi}\,\epsilon^{ij}g_{ab}\,\left(A^a{}_0F^b{}_{ij}+\partial_j(A^a{}_iA^b{}_0)\right).
}{eq:ch5} 
The total Hamiltonian is then given as $\mathcal{H}_T=\mathcal{H}+u^a{}_\mu\phi_a{}^\mu$, where $u^a{}_\mu$ are some arbitrary multipliers 
and $\phi_a{}^\mu$ are primary constraints. 
Conservation of these primary constraints, $\dot{\phi}_a{}^\mu=\{\phi_a{}^\mu,\,\mathcal{H}_T\}\approx0$, leads to the following secondary constraint\footnote{The Poisson bracket has its canonical form, 
\eq{
\{A^a{}_\mu(\textbf{x}),\pi_b{}^\nu(\textbf{y})\}=\delta^a_b\,\delta_\mu^\nu\,\delta^2(\textbf{x}-\textbf{y})\,.
}{} },
	\begin{equation}
		\mathcal{K}_a\equiv-\frac{k}{4\pi}\,\epsilon^{ij}g_{ab}F^b{}_{ij}\approx0\,,\qquad
	\end{equation}
 Defining $\bar{\mathcal{K}}_a=\mathcal{K}_a-\mathcal{D}_i\phi_a{}^i$ 
and forming the Poisson brackets between these constraints, it turns out that $\phi_a{}^0$ and $\bar{\mathcal{K}}_a$ are first class which means they have 
weakly vanishing Poisson brackets with all constraints in the theory
and $\phi_a{}^i$ are second class constraints. First-class property is preserved under Poisson brackets, consequently they can generate gauge transformations.

The algorithm how to construct these canonical generators out of first class constraints are due to Castellani \cite{Castellani:1981us}.
In our case the canonical generator is,
	\begin{equation}
		 G[\varepsilon]=\int_\Sigma\extd^2x\left(\mathcal{D}_0\varepsilon^a\pi_a{}^0+\varepsilon^a\bar{\mathcal{K}}_a\right)\,.
\label{eq:app1}
	\end{equation}
It is easy to show that the following gauge transformations are generated on the phase space by the Poisson bracket operation $\delta_{\varepsilon} \bullet =\{\bullet,G[\varepsilon]\}$ as expected,
\bea
\delta A^a{}_\mu=D_\mu\varepsilon^a\,,\qquad\qquad
\delta \phi_a{}^\mu=-f_{ab}{}^c\varepsilon^b\phi_c{}^\mu\,.
\eea
The generator $G[\varepsilon]$ is not yet functionally differentiable which means its variation in the field space is not only proportional to the variation of the field but also to the 
variation of its derivative,
\eq{
		\delta  G[\varepsilon]=\textrm{regular} - \frac{k}{2\pi}\int_\Sigma\extd^2x\,\varepsilon^{ij}\,\partial_i\langle\,\varepsilon\,,\delta A_j\rangle\,.
}{Intro:DeltaG}
The first term is the bulk variation of the generator. In the second term which is a boundary term we have strongly imposed the second class constraint $\phi_a{}^i$.
This boundary term spoils functional differentiability of our generator. In order to 
fix this, one adds a suitable boundary term $\delta Q$ to the variation of the canonical generator such that this additional boundary term cancels out;
	\begin{equation}
		 \delta\tilde G[\varepsilon]= \delta G[\varepsilon]+\delta Q[\varepsilon]\,,
\label{eq:app2}
	\end{equation}
 with
\begin{equation}\label{charge}
		\delta Q[\varepsilon]=\frac{k}{2\pi}\oint_{\partial\Sigma}\!\!\extd\varphi\, \langle\varepsilon\,,\delta A_\varphi\rangle\,.
	\end{equation}
If the gauge transformation parameter $\varepsilon$ is field independent this expression is also integrable and we can easily obtain the charge.
However if the gauge transformation parameter is state dependent integrability is not guaranteed.

In the special case -- in which we are interested -- where the gauge transformation parameter depends linearly on fields and not on 
their derivatives, $\varepsilon^a=A^a{}_\mu\xi^\mu$ and $\tau^a=A^a{}_\mu\eta^\mu$, gauge transformations generate diffeomorphisms on-shell \cite{Witten:1988hc},
\bea\label{difgtn}
\delta_\xi A^a{}_\mu=\{A^a{}_\mu,G[\xi]\}=\mathcal{L}_\xi A{}_\mu+\xi^\nu F^a{}_{\mu\nu}\,,
\eea
where
\bea
\mathcal{L}_\xi A{}_\mu=\partial_\mu \xi \cdot A+\xi\cdot\partial A_\mu\,.
\eea
These generators satisfy the following Poisson algebra on-shell,
\bea\label{PoissonGen}
\{\tilde{G}[\xi],\tilde{G}[\eta]\}&=&{G}[\zeta]+\tfrac{1}{2}\left(\delta_\eta Q[\xi]-\delta_\xi Q[\eta]\right)\,.
\eea
where $\zeta=[\xi,\eta]=\xi\ldotp\partial\eta-\eta\ldotp\partial\xi$ and the varied charges on the right hand 
side are computed via \eqref{difgtn}.

\section{Chern-Simons formulation of pure gravity}\label{se:3}
The first order action of pure gravity (without cosmological constant) in three dimensions can be written as,
\eq{
S=\frac{k}{4\pi}\,\int_{\mathcal{M}}{\langle e\wedge(\extd\om +\,\om\wedge\om)\rangle}_{\textrm{\tiny L}}\,,
}{}
where ${\langle\;,\;\rangle}_{\textrm{\tiny L}}$ is the invariant bilinear form of SO(2,1):
\eq{{\langle J_a,J_b\rangle}_{\textrm{\tiny L}}=\eta_{ab}\,,}{LBF}
 and $e$ and $\om$ are the three dimensional  SO(2,1)-valued one-forms denoted as the vielbein (dreibein) and the spin connection,
\eq{
e=(e^a{}_\mu dx^\mu) J_a \, \quad\text{and}\quad \om=(\om^a{}_\mu dx^\mu) J_a\,\quad\text{with}\quad J_a\in\text{SO(2,1)}\,.
}{}

This action can be written as a 
Chern-Simons gauge theory with  ISO(2,1) gauge group.\footnote{For an early work on this gauge group as a WZNW model see \cite{Salomonson:1989fw}.}
We start with the following algebra,
\begin{gather}\label{iso(2,1)}
[J_a,J_b]=\epsilon_{abc}J^c,\qquad
[J_a,P_b]=\epsilon_{abc}P^c,\qquad
[P_a,P_b]=0\,.
\end{gather}
This algebra allows for the following non-degenerate, invariant bilinear form:
\bea\label{form}
{\langle J_a,P_b\rangle}_{\textrm{\tiny P}}=\tfrac{1}{2}\eta_{ab}\,,
\eea
where $\eta_{ab}=(-,+,+)$ and $\epsilon^{012}=1$.
Changing the basis as,
\begin{gather}
L_{-1}=J_0-J_1,\qquad L_0=J_2,\qquad L_1=J_0+J_1\,,\nn\\
M_{-1}=P_0-P_1,\qquad M_0=P_2,\qquad M_1=P_0+P_1\,,
\end{gather}
we have,
\begin{gather}
[L_m,L_n]=(m-n)L_{m+n}\,,\qquad
[L_m,M_n]=(m-n)M_{m+n}\,,
\end{gather}
 where $m,n$ are $0,\pm1$. The bilinear form \eqref{LBF} in this basis is, 
 \bea
{\langle L_m,M_n\rangle}_{\textrm{\tiny P}}=\tfrac{1}{2}\eta_{mn}= \left( \begin {array}{ccc} 0&0&-1\\ \noalign{\medskip}0
&1/2&0\\ \noalign{\medskip}-1&0&0
\end {array} \right)\,.
\eea
We can now write the Chern-Simons action for this algebra,
where the gauge field as a Lie algebra valued one form isو
\bea
A =e^aP_a+\omega^aJ_a=A^n_M M_n+A^n_L L_n\,.
\eea

\subsection{Flat boundary conditions and charges}
In order to study flat space holography we work directly in the flat spacetime and consider the null infinity $\scri^+$, as the asymptotic
boundary. One could alternatively start with AdS results, take the limit of infinite radius in a proper way and obtain the 
flat results \cite{Barnich:2006av,Barnich:2013yka}.

We propose the following boundary conditions on the gauge field which is consistent with the variational principle and equations of motion ($F=0$),
\bea
\begin{array}{ lllll }\label{flatbcs}
  A_L^{1}= d\varphi		&&&&   A_M^{1}=du  \\
  A_L^{0}=0 					&&&&   A_M^{0}=r\,d\varphi  \\
  A_L^{-1}= -\frac{1}{4}\mathcal{M}\,d\varphi					&&&&   A_M^{-1}=-\frac{1}{4}\mathcal{M}\,du+\frac{1}{2}\,dr-\frac{1}{2}\mathcal{N}\,d\varphi\,,
\end{array}
\eea
where $\mathcal{N}=\mathcal{L}+\frac{u}{2}\mathcal{M}'$. The arbitrarily free functions $\mathcal{M}(\varphi)$ and $\mathcal{L}(\varphi)$ 
encode physical properties of different states in the theory. The spacetime coordinates $u$, $r$ and $\varphi$ range over the intervals $(-\infty,+\infty)$, $(0,+\infty)$ and $(0,2\pi)$ respectively.
The corresponding spacetime metric takes the following form,
\bea\label{fltmetr1}
ds^2&=&-(e^0)^2+(e^1)^2+(e^2)^2={\langle A_M, A_M\rangle}_{\textrm{\tiny L}}\nn\\
&=&\mathcal{M}du^2-2dudr+2\mathcal{N}dud\varphi + r^2d\varphi^2\,,
\eea
which is the leading contribution to the boundary conditions being used in the metric formalism \cite{Barnich:2006av,Barnich:2010eb}.
Minkowski background in Eddington--Finkelstein coordinate corresponds to
$\mathcal M = -1$ and $\mathcal N = 0$. In these coordinates $u$ plays the role of the light-like time at $\scri^+$.

Defining the parameters of translation and Lorentz transformation as, $\rho^a$ and $\tau^a$ respectively,
we can introduce a general gauge parameter w.r.t. the gauge group as $\varepsilon=\rho^aP_a+\tau^aJ_a=\epsilon_M^nM_n+\epsilon_L^nL_n$.
Then, the boundary-condition preserving gauge transformations (BCPGTs) of \eqref{flatbcs} are,
\bea
\begin{array}{ lllll }\label{fltBPST}
\epsilon_L^{1}=\epsilon							&&&&\epsilon_M^{1}=2\tau\\
\epsilon_L^{0}=-\epsilon' 						&&&&\epsilon_M^{0}=r\epsilon-2\tau'\\
\epsilon_L^{-1}=\frac{1}{2}\epsilon''-\frac{1}{4}\mathcal{M}\epsilon	&&&&\epsilon_M^{-1}=-\frac{1}{2}r\epsilon'+\tau''-\frac{1}{2}\mathcal{M}\tau -\frac{1}{2}\epsilon\mathcal{N}\,,
\end{array}
\eea
where $\tau=\sigma +\frac{u}{2}\epsilon'$, with $\sigma(\varphi)$ and $\epsilon(\varphi)$ being two arbitrary parameters.
We can read variations of the state dependent functions in \eqref{flatbcs} with respect to these parameters,
\bea\label{varflt1}
\delta_\epsilon \mathcal{L}&=& \epsilon\mathcal{L}'+2\epsilon'\mathcal{L}\,,\nn\\
\delta_\epsilon \mathcal{M}&=& \epsilon\mathcal{M}'+2\epsilon'\mathcal{M}-2\epsilon'''\,,\nn\\
\delta_\sigma \mathcal{L}&=& \sigma\mathcal{M}'+2\sigma'\mathcal{M}-2\sigma'''\,.
\eea

The varied boundary charge \eqref{charge} in this case is,
\bea\label{charge3}
 \delta Q[\varepsilon]=\frac{k}{2\pi}\int d\varphi\, {\langle\, \varepsilon  ,\delta A_\varphi\rangle}_{\textrm{\tiny P}} =\frac{k}{4\pi}\int d\varphi\, \, \left(\rho^a  \delta \omega_a{}_\varphi + \tau^a 
  \delta e_a{}_\varphi\right)\,.
\eea
Putting \eqref{flatbcs} and \eqref{fltBPST} into \eqref{charge3} we can integrate and read the corresponding charge,
\bea\label{charge0}
Q=\frac{k}{4\pi}\int d\varphi\,[\epsilon(\varphi)\mathcal{L}(\varphi)+\sigma(\varphi)\mathcal{M}(\varphi)]\,.
\eea
We can identify the vector fields in \eqref{PoissonGen} from the relation $\epsilon^a=A^a{}_\mu\xi^\mu$,
as $\xi^\mu=(2\tau,\epsilon,-r\epsilon')$. Writing them in terms of Fourier modes of parameters,
\bea\label{genflt}
\xi_n(\epsilon)=\xi(\epsilon=e^{in\varphi},\sigma=0)\qquad\text{and}\qquad
\xi_n(\sigma)=\xi(\sigma=e^{in\varphi},\epsilon=0)\,,
\eea
we can simply compute their nonzero Lie brackets,
\eq{
[\xi_n(\epsilon),\xi_m(\epsilon)]=i(m-n)\xi_{m+n}(\epsilon)\qquad\text{and}\qquad
[\xi_n(\epsilon),\xi_m(\sigma)]=i(m-n)\xi_{m+n}(\sigma)\,.
}{}
Using the identity \eqref{PoissonGen} with
variations on fields given in \eqref{varflt1} we find,
\bea
\tfrac{1}{2}\left(\delta_m Q_n[\epsilon]-\delta_n Q_m[\epsilon]\right)&=&i(m-n)Q_{m+n}[\epsilon]\nn\\
\tfrac{1}{2}\left(\delta_m Q_n[\epsilon]-\delta_n Q_m[\sigma]\right)&=&i(m-n)Q_{m+n}[\sigma]-i\,k\,n^3\delta_{m+n,0}\,,
\eea
and the Poisson bracket between these generators \eqref{PoissonGen} is found to be,
\begin{align}
i\{\tilde G_n(\epsilon),\,\tilde G_m(\epsilon)\} &= (n-m)\, \tilde G_{n+m}(\epsilon)\, ,\nn\\
i\{\tilde G_n(\epsilon),\,\tilde G_m(\sigma)\} &= (n-m)\, \tilde G_{n+m}(\sigma) + k\,n^3\,\de_{n+m,0}\, .
\end{align}
If we use the notation $L_n:=\tilde G_n(\epsilon)$ and $M_n:=\tilde G_n(\sigma)$ and
convert the Poisson brackets into quantum commutators 
by the prescription $i\{q,\,p\}=[\hat q,\,\hat p]$ -- and drop the hat for simplicity,
the non-zero commutators form a centrally extended BMS algebra in three dimensions (BMS$_3$),
\begin{align}\label{BMS1}
 [L_n,\,L_m] &= (n-m)\,L_{n+m}  +\frac{c_{\tiny L}}{12}\,(n^3-n)\,\de_{n+m,0}\, ,\nn\\
 [L_n,\,M_m] &= (n-m)\, M_{n+m} +\frac{c_{\small M}}{12}\,(n^3-n)\,\de_{n+m,0}\, .
\end{align}
In this case $c_{\small L}=0 $ and $c_{\tiny{M}} =12k$ and we have shifted the zero mode as, $M_0\to M_0+\frac{k}{2}$. 
This result for the ASA and its central charges, being found in CS formulation here, is in agreement with the result of \cite{Barnich:2006av,Barnich:2010eb} in metric formalism.
\section{Chern-Simons formulation of conformal gravity}\label{se:4}
The first order formulation of conformal gravity in three dimensions can be written in terms of three canonical variables,
\bea\label{1stcnf}
S=\frac{k}{4\pi}\int_{\mathcal{M}} {\langle\omega\wedge\left( d\omega 
+\tfrac{2}{3}\omega\wedge\omega\right)-2\lambda\wedge T\rangle}_{\textrm{\tiny L}} \,,
\eea
where again ${\langle\;,\;\rangle}_{\textrm{\tiny L}}$ is the same as before
and $\lambda$ is a three dimensional  SO(2,1)-valued one-form playing the role of a Lagrange multiplier to ensure 
the torsion constraint, $T=de+e\wedge\om=0$, such that in the end all quantities depend only on the dreibein.
Horne and Witten first wrote the Chern-Simons formulation of this theory based on SO(3,2) gauge group \cite{Horne:1988jf}
with the following algebra,
\begin{align}\label{so3,2}
[P_a,J_b]&=\epsilon_{abc}P^c,\;\quad\qquad[J_a,J_b]=\epsilon_{abc}J^c,\;\;\quad\qquad[K_a,J_b]=\epsilon_{abc}K^c,\nn\\
[P_a,D]&=P_a,\quad\quad[P_a,K_b]=-\epsilon_{abc}J^c+\eta_{ab}D,\quad\quad[K_a,D]=-K_a\,,
\end{align}
with $P_a$, $J_a$, $K_a$ and $D$ being generators of translation,
Lorentz transformation, special conformal transformations and dilatation, respectively.
The corresponding invariant bilinear form which is equivalent to the Killing form in this case is,
\eq{
{\langle J_a,J_b\rangle}_{\textrm{\tiny C}}=\eta_{ab},\qquad{\langle P_a,K_b\rangle}_{\textrm{\tiny C}}=-\eta_{ab},\qquad{\langle D,D\rangle}_{\textrm{\tiny C}}=1\,.
}{SO(3,2)Killing}
where the subscript {\tiny C} stands for the conformal group.
The gauge field as a Lie algebra-valued one form can be represented as,
\bea\label{connection1}
A_\mu dx^\mu=\left({e^a}_\mu P_a+{\omega^a}_\mu J_a+{\lambda^a}_\mu K_a+\phi_\mu D\right)dx^\mu\,.
\eea
Rewriting the Chern-Simons action \eqref{eq:ch1} 
in terms of these variables by using \eqref{SO(3,2)Killing} we find,
\bea\label{confphi}
I=-\frac{k}{4\pi}\int_{\mathcal{M}} \langle 2e\wedge d\lambda - \omega\wedge\left( d\omega 
+\tfrac{2}{3}\omega\wedge\omega \right)-\phi\wedge d\phi 
 +2e\wedge\left(\phi\wedge\lambda +\omega\wedge\lambda\right)\rangle_{\textrm{\tiny L}}\,,\nn\\
\eea
which is gauge equivalent to \eqref{1stcnf} (up to a boundary term). Indeed, below we will show that this corresponds to a 
gauge choice and $\phi$ is a Stu\"eckelberg field under the special conformal transformation (SCT).

A general Lie algebra-valued generator of gauge transformations in this case, can be written as,
\bea\label{gaugeCoNF}
\varepsilon=\rho^aP_a+\tau^aJ_a+\sigma^aK_a+\gamma D.
\eea
We may separate the state dependent and independent part of the gauge parameter as,
\bea\label{gtn}
  \begin{split}
    \rho^a &= e^a{}_\mu\xi^\mu +t^a\\
    \sigma^a &= \lambda^a{}_\mu\xi^\mu +s^a
  \end{split}
\qquad\qquad
  \begin{split}
    \tau^a &= \omega^a{}_\mu\xi^\mu +\theta^a\\
    \gamma &= \phi_\mu\xi^\mu +\Omega\,,
  \end{split}
\eea
where $t^a$ is just a global translation and can be absorbed in $\xi$, so we can put $t^a=0$.
The resulting on-shell gauge transformations are as follows,
\bea
\delta_\varepsilon e^a{}_\mu &=&\delta_\xi e^a{}_\mu +\epsilon^a{}_{bc}e^b{}_\mu\theta^c+\Omega e^a{}_\mu\,,\label{Vielbein}\\
\delta_\varepsilon \omega^a{}_\mu &=&\delta_\xi \omega^a{}_\mu +\mathcal{D}_\mu \theta^a-\epsilon^a{}_{bc}e^b{}_\mu s^c\,,\\
\delta_\varepsilon \lambda^a{}_\mu &=&\delta_\xi \lambda^a{}_\mu +\mathcal{D}_\mu s^a+\epsilon^a{}_{bc}\lambda^b{}_\mu\theta^c-\Omega \lambda^a{}_\mu +\phi_\mu s^a\,,\\
\delta_\varepsilon \phi_\mu &=&\delta_\xi \phi_\mu +\partial_\mu \Omega +\eta_{ab}e^a{}_\mu s^b\,.
\eea
The last identity shows that the gauge choice $\phi_\mu=0$ is accessible provided that,
\bea\label{gaugechoice}
s^a=-e^a{}^\nu\partial_\nu \Omega\,.
\eea
This gauge choice needs the invertibility of the vielbein and because \eqref{Vielbein} is unaffected by 
the gauge transformation \eqref{gaugechoice}, it remains invertible \cite{Horne:1988jf}. 

In the following we consider \eqref{confphi} in which the field $\phi$ and the gauge 
transformations associated to SCT are present, and base our analysis of conformal gravity on that.

\subsection{AdS boundary conditions and charges}
The fact that SO(3,2) contains SO(2,2) as a subgroup, suggests that we can study 
AdS boundary conditions in this setup. We should emphasize though, that the Chern-Simons theory obtained from the SO(2,2) subgroup of SO(3,2) is parity--odd. 
Although it is equivalent to the normal parity--even pure gravity with negative cosmological constant  at the level of equations of motion,  it is not equivalent to it at the level of action,
but to its ``exotic'' parity--odd partner \cite{Achucarro:1986vz,Witten:1988hc,Townsend:2013ela}. The normal one is a $\text{SL}(2,1)_k\oplus\text{SL}(2,1)_{-k}$ Chern-Simons theory while the exotic one
is a $\text{SL(2,1)}_k\oplus\text{SL(2,1)}_{k}$ theory. 
This is because the bilinear form of the exotic theory is induced from SO(3,2) Cartan-Killing form \eqref{SO(3,2)Killing}.

In the following we propose the AdS boundary conditions on the gauge field which is equivalent to those boundary conditions used in \cite{Afshar:2011qw} in metric formulation in 
the absence of partial massless mode -- for the case where this mode is turned on see subsection \ref{PMsec},
\bea\label{confbc1}
  \begin{split}
    e^0 &= -\ell e^{f}\left[e^{\rho}dt+e^{-\rho}\left(T_1dt -T_2d\varphi\right)\right]\,,\\
    e^1 &= -\ell e^{f}\left[e^{\rho}d\varphi-e^{-\rho}\left(T_1d\varphi -T_2dt\right)\right]\,,\\
    e^2 &= -\ell e^{f}d\rho\,,\\
    \lambda^a&=-1/2\ell^{-2} e^{-2f}e^a\,,
  \end{split}
\quad
  \begin{split}
    \omega^0 &= e^{\rho}\,d\varphi +e^{-\rho}\left(T_1\,d\varphi -T_2\,dt\right)\,,\\
    \omega^1 &= e^{\rho}\,dt-e^{-\rho}\left(T_1\,dt-T_2\,d\varphi\right)\,,\\
    \omega^2 &= 0\,,\\
    \phi&=\extd f\,,
  \end{split}
\eea
where $T_1(t,\varphi)$, $T_2(t,\varphi)$ and the Weyl factor, $f(t,\varphi)$, are some state dependent functions which are allowed to vary and specify our boundary 
conditions. We have introduced AdS radius $\ell$ as an emergent length scale in \eqref{confbc1}. In the asymptotic analysis we  note that,
\bea
\delta A_\mu dx^\mu=\delta(\ell e^{f} \bar A_\mu dx^\mu)=\ell e^{f}\left(\delta f  \,\bar A_\mu +\delta \bar A_\mu\right)dx^\mu\,.
\eea
The connection \eqref{connection1} with \eqref{confbc1} boundary conditions only satisfies flatness condition ($F=0$) when,
\bea
\partial_tT_2+\partial_\varphi T_1=0,\qquad\qquad\partial_\varphi T_2+\partial_tT_1=0\,.
\eea
which can be fulfilled if $T_2=\tfrac{1}{2}\left(\mathcal{L}(x^+)-\bar{\mathcal{L}}(x^-)\right)$ and $T_1=-\tfrac{1}{2}\left(\mathcal{L}(x^+)+\bar{\mathcal{L}}(x^-)\right)$.
The corresponding spacetime metric takes the form,
\bea\label{metricAdS}
ds^2&=&-(e^0)^2+(e^1)^2+(e^2)^2\nn\\
&=&\ell^2 e^{2f}\left[d\rho^2-\left(e^{2\rho}+\mathcal{L}(x^+)\bar{\mathcal{L}}(x^-)e^{-2\rho}\right)dx^+dx^-+\left(\mathcal{L}(x^+)dx^{+2}+\bar{\mathcal{L}}(x^-)dx^{-2}\right)\right]\,,\nn\\
\eea
where $x^{\pm}=\frac{t}{\ell}\pm\varphi$. BCPGTs are obtained in terms of the arbitrary functions, $a_1(t,\varphi)$, $a_2(t,\varphi)$ and $\Omega(t,\varphi)$ which parameterize these transformations,
\bea\label{BPGTcon}
  \begin{split}
\rho^0&=\ell e^{f}\left(a_2e^{\rho}+a_4e^{-\rho}\right)\,,\\
\rho^1&=\ell e^{f}\left(a_1e^{\rho}+a_3e^{-\rho}\right)\,,\\
\rho^2&=-\ell e^{f}\partial_\varphi a_2\,,\\
\sigma^a&=-1/2\ell^{-2} e^{-2f}\rho^a\,,
  \end{split}
\qquad
  \begin{split}
\tau^0&=-a_1e^{\rho}+a_3e^{-\rho}\,,\\
\tau^1&=-a_2e^{\rho}+a_4e^{-\rho}\,,\\
\tau^2&=\partial_\varphi a_2\,,\\
\gamma&=\Omega\,.
  \end{split}
\eea
Here $a_3=T_2a_2-T_1a_1-\tfrac{1}{2}\partial_\varphi^2 a_1$ and $a_4=T_1a_2-T_2a_1+\tfrac{1}{2}\partial_\varphi^2 a_2$.
The functions $a_1$ and $a_2$ should satisfy,
\bea
\partial_ta_2-\partial_\varphi a_1=0,\qquad\qquad\partial_ta_1-\partial_\varphi a_2=0\,,
\eea
which means $a_2=-\tfrac{1}{2}\left(\mathcal{\epsilon}(x^+)+\bar{\mathcal{\epsilon}}(x^-)\right)$ and $a_1=-\tfrac{1}{2}\left(\mathcal{\epsilon}(x^+)-\bar{\mathcal{\epsilon}}(x^-)\right)$.
Consistency of the variational principle \eqref{varprin} in the case where $\delta f\neq0$, gives the following condition,
\bea\label{varprWeyl}
\partial_\varphi f\,\partial_t \delta f-\partial_t f\,\partial_\varphi \delta f=0\,.
\eea
The following relation between the Weyl factor and the corresponding Weyl transformation is then dictated by the gauge invariance \eqref{deltI},
\bea\label{gaugeinvW}
\partial_\varphi f\partial_t \Omega -\partial_t f\partial_\varphi \Omega=0\,,
\eea
which is exactly the same relation in \cite{Afshar:2011qw} that guarantees the conservation of the Weyl charge. We satisfy these conditions by assuming,
\eq{f(t,\varphi)=f(x^+)\qquad\text{and}\qquad \Omega(t,\varphi)=\Omega(x^+) \,.}{}
We can read different variations,
\bea
\delta_{\epsilon} \mathcal{L}&=& \epsilon\mathcal{L}'+2\epsilon'\mathcal{L}-\frac{1}{2}\epsilon'''\,,\nn\\
\delta_{\bar{\epsilon}}\bar{\mathcal{L}}&=& \bar{\epsilon}\bar{\mathcal{L}}'+2\bar{\epsilon}'\bar{\mathcal{L}}-\frac{1}{2}\bar{\epsilon}'''\nn\,,\\
\delta_{\Omega} f&=& \Omega\,.
\eea

The varied boundary charge \eqref{charge} for this theory is obtained by inserting \eqref{connection1} and \eqref{gaugeCoNF},
\bea\label{charge1}
 \delta Q[\varepsilon]&=&\frac{k}{2\pi}\int d\varphi\, {\langle\, \varepsilon  ,\delta A_\varphi\rangle}_{\textrm{\tiny C}} \nn\\
  &=&-\frac{k}{2\pi}\int d\varphi\, \, {\langle \rho  \delta \lambda_\varphi - \tau \delta \omega_\varphi +\sigma 
  \delta e_\varphi -\gamma \delta \phi\rangle}_{\textrm{\tiny L}}\,.
\eea
At this stage we can show how this formula for the charge leads to the correct result in \cite{Afshar:2011qw}.
Using the gauge choice $\phi_\mu=0$ and plugging \eqref{gtn} and \eqref{gaugechoice} into the charge \eqref{charge1} we find,
\bea
\delta Q_P[\xi^{\mu}]&=&\frac{k}{2\pi}\int d\varphi\, \left[\xi^{\mu}\left(-e^a{}_{\mu}  \delta \lambda_a{}_\varphi +\omega^a{}_{\mu}  \delta \omega_a{}_\varphi -\lambda^a{}_{\mu}  \delta e_a{}_\varphi\right)+\theta^a\delta \omega_a{}_\varphi  \right]\nn\\
\delta Q_W[\Omega]&=&\frac{k}{2\pi}\int  d\varphi \,\left(e^a{}^\nu\partial_\nu \Omega\right) \delta e_a{}_\varphi\,
\eea
which matches precisely to the formula of charges obtained in \cite{Afshar:2011qw} upon a proper identification.

Plugging boundary conditions \eqref{confbc1} and their BCPGTs \eqref{BPGTcon} into \eqref{charge1} leads to,
\bea
 Q=\frac{k}{2\pi}\int d\varphi\, [\epsilon(x^+) \mathcal{L}(x^+)-\bar{\epsilon}(x^-)\bar{\mathcal{L}}(x^-)+\Omega(x^+)\partial_\varphi  f(x^+)]
\eea

After identifying $\xi$ and $\eta$ in \eqref{PoissonGen} from \eqref{confbc1} and \eqref{BPGTcon} and 
defining generators of these gauge transformations as,
\bea\label{genconf}
L_n=\tilde G[\epsilon=e^{inx^+}],\quad\quad
\bar L_n=\tilde G[\bar\epsilon=e^{inx^-}]\quad\text{and}\quad J_n=\tilde G[\Omega=e^{inx^+}]\,,
\eea
we can compute the Poisson bracket \eqref{PoissonGen}. As usual, we convert Poisson brackets into commutators by the prescription $i\{q,\,p\}=[\hat q,\,\hat p]$ -- and drop the hat for simplicity.
The resulting algebra is $\text{Vir}\oplus\overline {\text{Vir}}\oplus \hat {u}(1)_{k}$ with $c=-\bar c=6\,k$. 
We can now Sugawara-shift the quantum $L$ generator according to,
 \eq{
 L_m\to L_m + \frac{1}{2k} N(JJ)_{m},\qquad\quad N(JJ)_{m}=\,\sum_{n\in\mathbb{Z}} :J_n J_{m-n}:\, .
 }{eq:Lsug}
The non-zero commutators are,
\begin{align}\label{finalg1}
 [L_n,\,L_m] &= (n-m)\,L_{n+m} + \frac{c+1}{12}\,(n^3-n)\,\de_{n+m,0} \, ,\nn\\
 [\bar L_n,\,\bar L_m] &= (n-m)\,\bar L_{n+m} + \frac{\bar c}{12}\,(n^3-n)\,\de_{n+m,0}\, ,\nn\\
 [L_n,\,J_m] &= -m\,J_{n+m}\, ,\nn\\
 [J_n,\,J_m] &= k\,n\,\de_{n+m,0} \, ,
\end{align}
which shows a quantum shift by one in the central charge of one copy of Virasoro algebra. This is due to the normal ordering of $J$'s introduced in \eqref{eq:Lsug}.
There are potentially two interesting points in values of these central terms; the one corresponding to $k=-\tfrac{1}{6}$ which 
leads to a chiral half of Virasoro algebra with $\bar c=-1$ (the sign can be flipped by $\bar {L}_n\to -\bar{L}_{-n}$) and the one corresponding to $k=-\tfrac{1}{12}$ which leads to 
two copies of Virasoro algebra with the same central charge, $c+1=\bar c=\frac{1}{2}$.

\subsection{Partial massless modes in AdS}\label{PMsec}
The massless graviton modes $\mathcal {L}(x^+)$, $\bar{\mathcal {L}}(x^-)$ and the Weyl mode $f(x^+)$ are not the only modes 
appearing in conformal gravity on AdS. We can further turn on  partial massless (PM) modes \cite{Deser:1983tm,Deser:2001pe,Grumiller:2010tj,Afshar:2011qw}. 
The spin-2 PM  field  has one fewer degrees of freedom than the generic massive one. The massive graviton in three dimensions has 
two degrees of freedom, however in our context because of parity-odd nature of 3D conformal theory this is already reduced to one.
So all physical degrees of freedom of the PM mode remain at the boundary.

For simplicity in the following we put $f=0$ and $\ell=1$, this does not reduce the generality of our analysis and we can recover
it in the end. If we define the following PM one forms,
\bea\label{PM-1-forms}
  \begin{split}
    p^0 &=  P_2 dt -P_1 d\varphi\,,
  \end{split}
\quad
  \begin{split}
    p^1 &= P_1 dt -P_2 d\varphi\,,
  \end{split}
  \quad
  \begin{split}
    p^2 &= 0\,,
  \end{split}
\eea
we can write the new PM contribution to our connection \eqref{connection1} as,
\bea\label{PM-1-forms1}
  \begin{split}
    e^a &\to e^a + p^a\,,
  \end{split}
\qquad
  \begin{split}
    \lambda^a &\to \lambda^a +\tfrac{1}{2}\, p^a\,.
  \end{split}
\eea
Solving equations of motion,  the following restrictions on modes are found,
\begin{gather}
    \partial_tP_2+\partial_\varphi P_1=0,\qquad\qquad\partial_\varphi P_2+\partial_tP_1=0\,,\label{PMcon1}\\ 
    P_1^2 -P_2 ^2=0\,,\qquad\qquad   T_1P_1-T_2P_2=0\,.\label{PMcon2}
\end{gather}
The familiar equation \eqref{PMcon1} just suggests to write $P_2=\tfrac{1}{2}\left(\mathcal{P}(x^+)-\bar{\mathcal{P}}(x^-)\right)$ and $P_1=-\tfrac{1}{2}\left(\mathcal{P}(x^+)+\bar{\mathcal{P}}(x^-)\right)$, 
while the equation \eqref{PMcon2} necessitates to kill one sector of the PM modes as well as one sector of the massless 
ones \footnote{In fact if we relax the last boundary condition in \eqref{PM-1-forms} (weighted by $e^{-\rho}$), one can keep both sectors \cite{Afshar:2011yh,Afshar:2011qw}.},
\eq{ \bar{\mathcal{P}}(x^-)=0\,,\qquad\qquad \bar{\mathcal{L}}(x^-)=0\,.}{chiralPMcon}
The metric \eqref{metricAdS} changes to,
\bea\label{metricPM}
ds^2=\ell^2 e^{2f}\left[d\rho^2-e^{2\rho}dx^+dx^-+\left(\mathcal{P}(x^+)e^\rho +\mathcal{L}(x^+)\right)dx^{+2}\right]\,.
\eea

As a consequence of \eqref{chiralPMcon}, the bar sector of the gauge transformations also becomes zero, $\bar{\mathcal{\epsilon}}(x^-)=0$.
BCPGTs in this case can be obtained from \eqref{BPGTcon} by,
\bea\label{PM-BGT}
  \begin{split}
    \rho^0 &\to \rho^0 +\tfrac{1}{2}\, \epsilon(x^+)\mathcal{P}(x^+)\,,\\
    \sigma^0 &\to \sigma^0 +\tfrac{1}{4}\, \epsilon(x^+)\mathcal{P}(x^+)\,,
  \end{split}
\qquad
  \begin{split}
    \rho^1 &\to \rho^1 -\tfrac{1}{2}\, \epsilon(x^+)\mathcal{P}(x^+)\,.\\
    \sigma^1 &\to \sigma^1 -\tfrac{1}{4}\, \epsilon(x^+)\mathcal{P}(x^+)\,.
  \end{split}
\eea
and putting $\bar{\mathcal{\epsilon}}(x^-)=0$. These gauge transformations induce the following variations on fields,
\bea
\delta_{\epsilon} \mathcal{L}&=& \epsilon\mathcal{L}'+2\epsilon'\mathcal{L}-\frac{1}{2}\epsilon'''\,,\nn\\
\delta_{\epsilon} \mathcal{P}&=& \epsilon\mathcal{P}'+\frac{3}{2}\epsilon'\mathcal{P}\,.
\eea
So the presence of the PM mode leads to a chiral half of Virasoro algebra with $c=6\,k$ and a conformal--dimension--$\frac{3}{2}$ current.
Note that although the boundary PM mode plays the role of a current w.r.t. the Virasoro generator, it is not part of the ASA. 
We can also observe this by looking into the commutators of zero modes of the PM generator \eqref{PMalgL} with sl(2,R) generators,
\eq{
[L_1,X_-]=X_2+D\,,\qquad [L_0,X_-]=\tfrac{1}{2}X_-\,,\qquad[L_{-1},X_-]=0\,.
}{PMalg}
Since in our boundary conditions the PM mode is only coupled to $X_-$, and not to $X_2$ and $D$ -- which is necessary to solve $F=0$ equations --
the Jacobi identity is not satisfied.

In the presence of the Weyl factor, $f(x^+)\neq0$, we are led to the same conclusion as in \eqref{finalg1}; adding the $\hat{u}(1)_k$ current and shifting the central charge by one.

\subsection{Flat boundary conditions and charges}

We can also impose flat boundary conditions in the conformal Chern-Simons theory by restricting ourselves to the ISO(2,1)
subgroup of the conformal group \eqref{so3,2}. The nice feature in this case is the role played by the 
bilinear form inherited from the SO(3,2) Killing form. It obviously keeps the theory parity-odd and as we will see this leads to switching the central term 
in the algebra \eqref{BMS1} and introduces a chiral half of Virasoro \cite{Bagchi:2012yk}. 

The imposed flat boundary conditions in this case would be essentially the same as \eqref{flatbcs} for the  ISO(2,1) part with the same BCPGTs \eqref{fltBPST} plus 
additional conditions on other fields,

\bea\label{confbc6}
  \begin{split}
    e^0 &=  e^{f}\left[\left(1 -\tfrac{1}{4}\mathcal{M}\right)dt-\tfrac{1}{2}\mathcal{N}d\varphi +\tfrac{1}{2}dr\right]\,,\\
    e^1 &=  e^{f}\left[\left(1 +\tfrac{1}{4}\mathcal{M}\right)dt+\tfrac{1}{2}\mathcal{N}d\varphi -\tfrac{1}{2}dr\right]\,,\\
    e^2 &=  e^{f}\,r\,d\varphi\,,\\
    \lambda^a&=0\,,
  \end{split}
\quad
  \begin{split}
    \omega^0 &= \left(1 -\tfrac{1}{4}\mathcal{M}\right)d\varphi\,,\\
    \omega^1 &= \left(1 +\tfrac{1}{4}\mathcal{M}\right)d\varphi\,,\\
    \omega^2 &= 0\,,\\
    \phi&=\extd f\,,
  \end{split}
\eea
and 
\bea\label{BPfltcon}
  \begin{split}
\rho^0&= e^{f}\left[2\left(1-\tfrac{1}{4}\mathcal{M}\right)\tau-\tfrac{1}{2}r\epsilon'+\tau''-\tfrac{1}{2}\epsilon\mathcal{N}\right]\,,\\
\rho^1&= e^{f}\left[2\left(1+\tfrac{1}{4}\mathcal{M}\right)\tau+\tfrac{1}{2}r\epsilon'-\tau''+\tfrac{1}{2}\epsilon\mathcal{N}\right]\,,\\
\rho^2&= e^{f}\left(r\epsilon-2\tau'\right)\,,\\
\sigma^a&=0\,,
  \end{split}
\qquad
  \begin{split}
\tau^0&=\left(1-\tfrac{1}{4}\mathcal{M}\right)\epsilon +\tfrac{1}{2}\epsilon''\,,\\
\tau^1&=\left(1+\tfrac{1}{4}\mathcal{M}\right)\epsilon -\tfrac{1}{2}\epsilon''\,,\\
\tau^2&=-\epsilon'\,,\\
\gamma&=\Omega\,.
  \end{split}
\eea
The spacetime metric is conformal to the flat metric in BMS gauge,
\bea\label{fltmetr2}
ds^2&=&-(e^0)^2+(e^1)^2+(e^2)^2={\langle e, e\rangle}_{\textrm{\tiny L}}\nn\\
&=& e^{2f}\left[\mathcal{M}du^2-2dudr+2\mathcal{N}dud\varphi + r^2d\varphi^2\right]\,,
\eea
where $\mathcal{N}=\mathcal{L}+\frac{u}{2}\mathcal{M}'$. 
The Weyl factor and its variation (like other functions) depend only on $\varphi$ to guarantee the well-defined variational principle \eqref{varprWeyl},
gauge invariance \eqref{gaugeinvW} and also conservation of the Weyl charge.
Different variations of state dependent variables read as,
\bea\label{fltCSvar}
\delta_\epsilon \mathcal{L}&=& \epsilon\mathcal{L}'+2\epsilon'\mathcal{L}\,,\nn\\
\delta_\epsilon \mathcal{M}&=& \epsilon\mathcal{M}'+2\epsilon'\mathcal{M}-2\epsilon'''\,,\nn\\
\delta_\sigma \mathcal{L}&=& \sigma\mathcal{M}'+2\sigma'\mathcal{M}-2\sigma'''\,,\nn\\
\delta_{{\tiny \Omega}}f&=& \Omega\,.
\eea
The boundary charge in this case can be read from \eqref{charge1} using \eqref{confbc6} and \eqref{BPfltcon},
\bea\label{charge2}
Q=\frac{k}{2\pi}\int d\varphi\,[\tfrac{1}{2}\epsilon(\varphi)\mathcal{M}(\varphi)+\partial_\varphi f(\varphi)\Omega(\varphi)]\,.
\eea
Although the $\mathcal{L}$ and $\mathcal{M}$ sector of variations in \eqref{fltCSvar} are exactly the same as 
\eqref{varflt1}, the corresponding charges \eqref{charge0} and \eqref{charge2} are different; 
the response function $\mathcal{M}(\varphi)$ in \eqref{charge2} is now coupled to $\epsilon$ instead of $\sigma$.

Using the identity \eqref{PoissonGen} and variations given in \eqref{fltCSvar} and writing generators 
in terms of Fourier modes as in \eqref{genflt}, we are led to a centrally extended algebra, $\text{BMS$_3$}\oplus \hat {u}(1)_{k}$
with $c_L=12\,k$ and $c_M=0$ as the asymptotic symmetry algebra. 
Such  an ASA  without Weyl rescaling charges was also found in \cite{Bagchi:2012yk} as a scaling limit of topologically massive gravity (TMG) result, in metric formulation.
Sugawara-shifting Virasoro generators as in \eqref{eq:Lsug} leads to the same  conclusion as before; quantum shifting 
the  central charge in one copy of the Virasoro algebra by unity, $c_L\to c_L+1$, and adopting $J$'s as $\hat {u}(1)_k$ currents w.r.t. Virasoro generators,
\begin{align}\label{Vir-M}
[L_n,L_m]&=(n-m)L_{m+n}+\frac{c_L+1}{12}\,n(n^2-1)\delta_{m+n,0}\,,\nn\\
[L_n,M_m]&=(n-m)M_{m+n}\,,\nn\\
[L_n,J_m]&=-mJ_{m+n}\,,\nn\\
[J_n,J_m]&=k\,n\,\delta_{m+n,0}\,.
\end{align}
The Sugawara-construction analogue of \eqref{eq:Lsug} in supertranslation generators, $M$ is in principle also possible. That makes their commutator anomalous,
which is intuitively in contrast with their role as infinite extensions of translation generators.

\subsection{Representation of the ASA}
Finally let us discuss the representation of the algebra \eqref{Vir-M}. Due to the presence of an additional $\hat{u}(1)_k$ current there 
are now three quantum numbers associated to $J_0$, $L_0$ and $M_0$ by which each state is labeled,
\begin{align}
L_0|h,\xi,q\rangle&=h|h,\xi,q\rangle\nn\\
M_0|h,\xi,q\rangle&=\xi|h,\xi,q\rangle\nn\\
J_0|h,\xi,q\rangle&=q|h,\xi,q\rangle\,,
% ...
\end{align}
where $\xi$ is the $M_0$ eigenvalue denoted as ``rapidity'' \cite{Bagchi:2009pe}.
We can define the highest weight state by demanding to be annihilated by all positive modes $J_n$, $L_n$ and $M_n$ where $n>0$. 
The action of negative--mode generators on these states create new descendants. 
Since we are dealing with flat Minkowski background, the vacuum of the theory should be iso(2,1) invariant which means $h=\xi=0$.
On the other hand, since $c_M=0$, we can decouple all $M_n$ descendants and reduce the tower of descendants to the Virasoro's 
and the $\hat {u}(1)_k$ current's \cite{Bagchi:2009pe}.
Unitarity requires positive norms for all states, so already at level one the norm of the state $J_{-1} |q\rangle$ gives $k>0$. 
At level $N=2$ there are three states appearing; $L_{-2} |q\rangle$, $J_{-2} |q\rangle$ and $J_{-1}^2 |q\rangle$. The Ka\v{c} determinant is,
\eqa{
K_2(q,k)
&=&2k^2\left(k\,c_L-4q^2\right)=8k^2(3k^2-q^2)\,,
}{}
where in the last equality we have set $c_L=12k$. Here we see that the unitarity bound is shifted to $c_L=4q^2/k$ at which we have one null state at level $N=2$.
At level $N=3$, there are five states appearing as descendants; $L_{-3} |q\rangle$, $J_{-1}L_{-2} |q\rangle$, $J_{-3} |q\rangle$, $J_{-1} J_{-2}|q\rangle$  and $J_{-1}^3 |q\rangle$. 
The Ka\v{c} determinant in this case is,
\eqa{
K_3(q,k)=18 k^5 (k\,c_L  - 4 q^2) (2k\, c_L  - 3 q^2)\,,
}{}
which shows a new null state for $c_L=3q^2/2k$ at level $N=3$. If we turn off the Weyl mode by setting all $J$'s to zero these quantities are respectively $\bar{K}_1=0$, $\bar{K}_2=c_L/2$ and $\bar{K}_3=2c_L$.
Thus, in the presence of the Weyl charge, the physical Hilbert space is reducible.
This is an example showing how adding bulk gauge symmetries (Weyl symmetry in this case) leads to removing some of the perturbative
states from the spectrum \cite{Castro:2011ui}. 

Similarly the representation of the ASA in \eqref{BMS1} and \eqref{finalg1} can be studied. In \eqref{BMS1}, where all Weyl charges are zero and $c_M\neq0$, the Ka\v{c} 
determinant at level $N=2$, is proportional to $-c_M^2$ which shows non-unitarity. In AdS case, \eqref{finalg1}, all results above apply by changing $c_L$ to $c$.

\section{Summary}\label{se:5}
In this paper we studied different boundary conditions in CS formulation of 3D pure gravity and conformal gravity.
In the pure gravity case, by imposing flat boundary conditions on the gauge field at the asymptotic null infinity, $\scri^+$, 
we recovered the famous asymptotic BMS$_3$ symmetry algebra, \cite{Barnich:2006av}. We also imposed the same set of boundary
conditions in the theory of SO(3,2) CS gauge theory which is known to be equivalent to 3D conformal gravity \cite{Horne:1988jf} and found the ASA.
The central charge in the ASA of this theory appears in the commutator 
of Virasoro generators opposite to the pure gravity case where it is in the commutator of supertranslation generators.
This was first shown in \cite{Bagchi:2012yk} as a flat space limit of the ASA of TMG and confirmed here by direct computation in conformal gravity.
The other difference is the presence of an additional $\hat{u}(1)_k$ current related to the dilatation gauge transformation and 
the quantum shift of the central charge. We also investigated the representation of this algebra and showed that the presence of 
the Weyl charge can lead to null boundary states.

We also considered two other sets of boundary conditions for the conformal theory. By imposing 
AdS boundary conditions we confirmed the result in \cite{Afshar:2011yh,Afshar:2011qw}, namely having two copies of Virasoro algebra with an additional $\hat{u}(1)_k$ current as ASA.
Relaxing boundary conditions in favor of partial massless mode kills one half of Virasoro generators and we end up with
one copy of Virasoro and an additional (PM) current with conformal dimension-$\frac{3}{2}$.

\acknowledgments
I am very grateful to Daniel Grumiller and Jan Rosseel for many useful discussions.
I thank Arjun Bagchi, Glenn Barnich, Reza Fareghbal and Niklas Johansson for their useful comments.
I also thank the Galileo Galilei Institute for Theoretical Physics for
the hospitality and the INFN for partial support during ``Higher Spins, Strings and Duality''
Workshop.
I was supported by the START project Y 435-N16 of the Austrian
Science Fund (FWF) and the FWF projects I 952-N16
and I 1030-N27.

\appendix
\section{Connections in the light-cone gauge}
In this appendix we make the connection to the usual light cone representation of generators which is used in 
AdS literature. First we present the algebra \eqref{so3,2} in a new basis,
\bea\label{so3,2-1}
  \begin{split}
[J_a,J_b]&=\epsilon_{abc}J^c\,,\\
[J_a,X_b]&=\epsilon_{abc}X^c\,,\\
[X_a,X_b]&={\Lambda}\epsilon_{abc}J^c\,,\\
[X_a,D]&=Y_a\,,
  \end{split}
\qquad
  \begin{split}
[X_a,Y_b]&=\Lambda\eta_{ab}D\,,\\
[J_a,Y_b]&=\epsilon_{abc}Y^c\,,\\
[Y_a,Y_b]&=-{\Lambda}\epsilon_{abc}J^c,\\
[Y_a,D]&=X_a\,,
  \end{split}
\eea
where $\eta=(-,+,+)$ and $\epsilon^{012}=1$. Here we have introduced $\Lambda$ as an emergent parameter and,
\bea
X_a=P_a-\tfrac{\Lambda}{2}K_a,\qquad\qquad Y_a=P_a+\tfrac{\Lambda}{2}K_a\,.
\eea
For $\Lambda<0$ we can identify $J_a$ and $ Y_a$ as  generators of so(2,2) subalgebra,
$J_a$, and $ X_a$ as generators of so(3,1) subalgebra of so(3,2) \eqref{so3,2-1}.
So $X_a$ and $Y_a$ are playing the role of translation in dS and AdS spaces respectively. 
For so(2,2) subalgebra, by introducing,
\bea
J^{L}_a=\frac{1}{2}\left(J_a+\frac{1}{\sqrt{-\Lambda}}Y_a\right)\,,\qquad J^{R}_a=\frac{1}{2}\left(J_a-\frac{1}{\sqrt{-\Lambda}}Y_a\right)\,,
\eea
we can simply show that
this algebra can be written as sl(2,R)$_L\oplus$ sl(2,R)$_R$,
where $J^{l/R}_a$ are generators of sl(2,R)$_{L/R}$.
Introducing $ L_{\pm1}= J^R_{\pm}$,  
$ L_0=J^R_2$ and $\bar{ L}_{\pm1}= J^L_{\pm}$, $\bar{ L}_0=J^L_2$ we have,
\begin{gather}
[ L_m, L_n]=(m-n) L_{m+n}\,,\qquad
[\bar{ L}_m,\bar{ L}_n]=(m-n)\bar{ L}_{m+n}\,,\qquad
[ L_m,\bar { L}_n]=0\,.
\end{gather}
where $m,n=\pm10$ and $J_{\pm}=J_0\pm J_1$. In this basis, $\eta_{+-}=-2\eta_{22}=-2$ and $\epsilon_{+-2}=2$. We can represent 
the remaining generators, $P_{-\frac{1}{2}}=\frac{1}{\sqrt{-\Lambda}}X_{-}$ and $P_{\frac{1}{2}}=\frac{1}{\sqrt{-\Lambda}}X_2+ \,D$ of the full so(3,2) 
algebra w.r.t. its sl(2,R) subalgebra,
\eq{
[L_n,P_m]=\left(\frac{n}{2}-m\right)P_{m+n}\,.}{PMalgL}
and the same for $\bar{P}_{-\frac{1}{2}}=\frac{1}{\sqrt{-\Lambda}}X_2- \,D$ and $\bar{P}_{\frac{1}{2}}=\frac{1}{\sqrt{-\Lambda}}X_{+}$.

Using the notation introduced above and considering $\Lambda=-\ell^{-2}e^{-2f}$, we can translate the most general form  of 
connection \eqref{connection1} which was used for  AdS boundary conditions in the light cone coordinate as,
\bea
A&=&\left(L_1\,dx^+-\bar{L}_{-1}\,dx^-\right)e^{\rho}+\left(L_0-\bar{L}_0\right)d\rho\qquad\quad\text{AdS background}\nn\\
&+&\frac{1}{2}\left(\mathcal{P}(x^+)P_{-\tfrac{1}{2}}\,dx^+-\bar{\mathcal{P}}(x^-)\bar{P}_{\tfrac{1}{2}}\,dx^-\right)\qquad\;\quad\text{Partial massless modes}\nn\\
&-&\left(\mathcal{L}(x^+)L_{-1}\,dx^+-\bar{\mathcal{L}}(x^-)\bar{L}_{1}\,dx^-\right)e^{-\rho}\qquad\;\;\;\quad\text{Massless modes}\nn\\
&+&\left(\partial_+f(x^+)\,dx^++\partial_-\bar{f}(x^-)\,dx^-\right)D\qquad\;\;\,\,\,\qquad\text{Weyl modes}\,.
\eea

%
\providecommand{\href}[2]{#2}\begingroup\raggedright\endgroup

\end{document}